\documentclass[useAMS]{mn2e}

\usepackage{times}
\usepackage{psfig}
\usepackage{graphicx}

\def\spose#1{\hbox to 0pt{#1\hss}}
\newcommand\lsim{\mathrel{\spose{\lower 3pt\hbox{$\mathchar"218$}}
     \raise 2.0pt\hbox{$\mathchar"13C$}}}
\newcommand\gsim{\mathrel{\spose{\lower 3pt\hbox{$\mathchar"218$}}
     \raise 2.0pt\hbox{$\mathchar"13E$}}}

\title[Intergalactic magnetic fields and GeV emission of 1ES 0229+200] 
{The intergalactic magnetic field constrained by {\it Fermi}/LAT observations of 
the TeV blazar 1ES 0229+200}

\author[Tavecchio et al.]
{F. Tavecchio$^1$\thanks{E--mail: fabrizio.tavecchio@brera.inaf.it}, 
G. Ghisellini$^1$, L. Foschini$^1$, G. Bonnoli$^1$, G. Ghirlanda$^1$, P. Coppi$^2$
\\
$^1$INAF -- Osservatorio Astronomico di Brera, via E. Bianchi 46, I--23807 Merate, Italy\\
$^2$ Dept. of Physics, Yale University, P.O. Box 208120, New Haven, CT 06520-8120, USA
}

\begin{document}


\pagerange{\pageref{firstpage}--\pageref{lastpage}} \pubyear{2008}

\maketitle

\label{firstpage}

\begin{abstract}
TeV photons from blazars at relatively large distances, interacting with the optical--IR cosmic 
background, are efficiently converted into electron--positron pairs. 
The produced pairs are extremely relativistic (Lorentz factors of the order of $10^6$--$10^7$) 
and promptly loose their energy through inverse Compton scatterings with the photons of the 
microwave cosmic background, producing emission in the GeV band. 
The spectrum and the flux level of this reprocessed emission is critically dependent 
on the intensity of the intergalactic magnetic field, $B$, that can deflect the pairs 
diluting the intrinsic emission over a large solid angle.
We derive a simple relation for the reprocessed spectrum expected from a steady source. 
We apply this treatment to the blazar 1ES 0229+200, whose intrinsic very hard 
TeV spectrum is expected to be approximately steady. 
Comparing the predicted reprocessed emission with the upper limits measured 
by the {\it Fermi}/Large Area Telescope, we constrain the value of the 
intergalactic magnetic field to be larger than $B \simeq 5\times 10^{-15}$ Gauss,
depending on the model of extragalactic background light.
\end{abstract}

\section{Introduction}

Although cosmic magnetic fields are present at all scales and 
likely play a non negligible role in shaping the Universe, our 
knowledge of them is still rather poor. In particular, magnetic fields at  
cosmological scales ($>1$ Mpc) remain elusive (e.g. Kronberg 2001, Neronov \& Semikoz 2009).
From the theoretical point of view, intergalactic space can contain the 
traces of magnetic fields produced in the initial phases of the Universe 
(e.g., Grasso \& Rubinstein 2001) and can also be polluted by magnetic 
field ejected by galaxies and quasars (e.g. Furlanetto \& Loeb 2001). 
Methods based on the rotation measure in the radio band constrain the 
intensity of the intergalactic magnetic field (IGMF) below $B<10^{-11}$--$10^{-9}$ G 
(e.g. Kronberg 2001 and references therein) but these values depends on the 
highly uncertain correlation length of the field.

As early proposed by Plaga (1995), IGMF could be probed by 
using indirect methods based on the effects of the IGMF on 
the electron--positron pairs produced through the interaction of a TeV photon from a 
cosmological source with a low energy photon of the optical--IR cosmic background. 
The produced pairs promptly loose their energy through inverse Compton (IC) 
scattering, in which a photon of the cosmic microwave background (CMB) 
is boosted at $\gamma$--ray (GeV) energies. 
The existence of a magnetic field, even with the tiny intensities suggested by 
the measures mentioned above, modifies the properties of the $\gamma$--ray emission, 
causing observable effects (time delays, e.g. Dai \& Lu 2002; formation of extended 
$\gamma$--ray halos, Aharonian, Coppi \& Voelk 1994, Neronov et al. 2010) 
that can be exploited to infer the value of $B$ (e.g. Neronov \& Semikoz 2009 
and references therein). 
The reprocessing of multi--TeV emission from blazars into GeV--MeV emission 
has been also proposed has a possible contribution to the observed $\gamma$--ray 
background (Coppi \& Aharonian 1997, Venters 2010). 

In this paper we use a method based on the observed level of the GeV emission of an 
highly absorbed BL Lac, 1ES 0229+200, whose intrinsic TeV spectrum is expected 
to be almost steady and very hard (Aharonian et al. 2007, Tavecchio et al. 2009). 
Several papers discussed a similar method applied to the emission from 
gamma--ray bursts (e.g. Dai \& Lu 2002, Razzaque et al. 2004, Takahashi et al. 2008). 
A similar method for blazars have been already outlined in Dai et al. (2002) 
and Murase et al. (2008), but these works were focused on the discussion of 
the {\it variable} GeV emission, by--product of the reprocessing of the primary 
TeV emission of a rapidly flaring blazar. 
Here, instead, we present a simplified treatment of the case of stationary emission 
of the primary TeV source, suitable to describe the case of 1ES 0229+200. 
Basically, our approach is based on the fact that for increasing values of the IGMF the level of expected reprocessed emission in the MeV-GeV band  decreases, since the total amount of the primary flux is spread over larger solid angles. The comparison between the expected level of the reprocessed flux and the flux at GeV energies observed by the {\it Fermi}/Large Area Telescope thus provide 
a value (or a lower limit) on the intensity of the IGMF.

We use the notation $Q=Q_X 10^X$ in cgs units.

After the completion of this work a paper on same argument have been published (Neronov \& Vovk 2010). The authors derive a value of the magnetic field very close to the value derived here, using a full treatment considering also the development of a cascade. 
However, they implicitly assume that the source is emitting TeV radiation 
{\it isotropically}. In this case the total flux of the reprocessed emission does not depends on 
the IGMF. The IGMF has the only effect to reduce the expected surface brightness of the 
reprocessed emission, lower for larger values of the magnetic field. Instead, in our work we 
assume that the primary high-energy photons from the blazar, as commonly assumed, are strongly 
beamed since they are produced in a narrow relativistic jet. 
In this case the effect of the IGMF 
is to spread the reprocessed emission in solid angles larger than the original beaming 
angle and thus reduce the total observed flux, even if the source remains unresolved.

\begin{figure}
\hskip -1.7truecm
\psfig{file=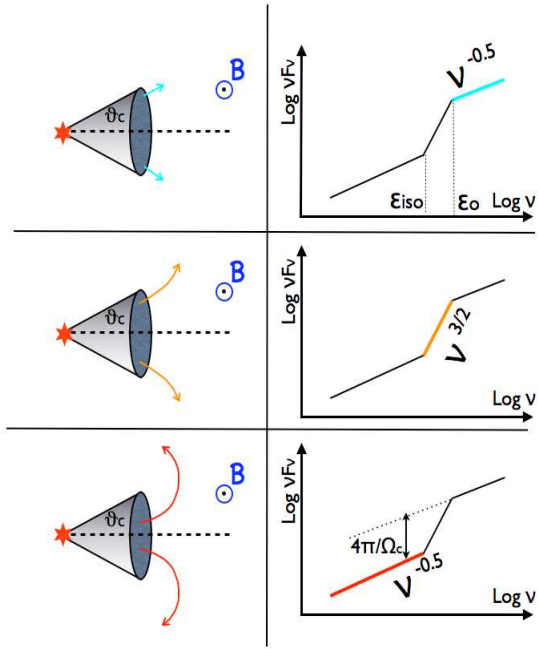,height=8.9cm,width=10.9cm}  
\vskip 0.5 true cm
\caption{
Cartoon of the reprocessing of the absorbed TeV radiation. 
The source illuminates the regions inside a cone with semi--aperture $\theta _{\rm c}$. 
TeV photons are converted into e$^{+/-}$ pairs at a typical distance of hundreds of Mpc. 
Pairs cool rapidly through inverse Compton scattering on the CMB. 
For a given intensity of the (perpendicular component of the) 
extragalactic magnetic field, pairs with very high energy (upper panel) 
cool without changing their direction (arrows) and thus the reprocessed GeV 
emission from the IC is beamed within the same angle $\theta _{\rm \,\, c}$. 
The emission of progressively lower energy of the pairs is instead spread 
over larger angles, due to the curved trajectories of the pairs, in turn due to
their longer lifetimes. 
As detailed in the text the resulting overall spectrum (right panels) can 
be approximates by three power laws, each branch due to electrons emitting 
within the original cone (upper), within a cone larger than the original 
(middle) and almost isotropically (lower).
}
\label{cartoon}
\end{figure}

\section{The reprocessed spectrum}

The interaction of a  $\gamma$--ray with energy $E= 1\, E_{\rm TeV}$ TeV  
with a low frequency (optical--IR) photon of the extragalactic background light 
(EBL) produces an electron positron pair with Lorentz factor 
$\gamma \simeq E/2m_e c^2\simeq 10^6 E_{\rm TeV}$ (for simplicity and since the redshift of 1ES 0229+200 is small, we neglect here and in the following the redshift dependence of the photon energies). 
These pairs, in turn, will inverse Compton scatter the photons 
of the CMB producing $\gamma$--rays of (observed) energy 
$\epsilon \simeq \gamma^2 h \nu_{\rm CMB} \simeq  2.8\, kT_{\rm CMB}\, \gamma^2 = 0.63\, E_{\rm TeV}^2$ GeV. 
In the following we assume that the maximum energy of the photons 
produced through IC scattering, $\epsilon_{\rm max}$, is below the 
threshold of absorption by the EBL (approximately above 400-500 GeV for $z=0.14$ for most of the EBL models). 
For higher $\epsilon_{\rm max}$ an electromagnetic cascade would develop and a 
full numerical treatment is needed (e.g. Coppi \& Aharonian 1997, d'Avezac et al. 2007; 
but see \S4 below).

The typical distance from the blazars at which $\gamma$--rays are absorbed 
is of the order of  few hundreds Mpc for 1--10 TeV photons  (e.g. Dermer 2007). 
On the other hand, once produced, pairs cool very rapidly, 
the cooling time for IC scattering being: 
\begin{equation}
t_{\rm cool}=\frac{3m_ec}{4\gamma \sigma _T U_{\rm CMB,0}(1+z_{\rm r})^4}\simeq 7\times 10^{13} \gamma _6^{-1} (1+z_{\rm r})^{-4} {\rm s}
\end{equation}
\noindent
translating into a distance $ct_{\rm cool}=2\times 10^{24} \gamma^{-1}_6 (1+z_{\rm r})^{-4}$ cm. 
Here $U_{\rm CMB,0}$ is the energy density of the cosmic background radiation evaluated
at $z=0$, and $z_{\rm r}$ is the redshift of the reprocessing region (not necessarily equal to 
the redshift of the primary source).

Therefore, at least for the pairs with the largest energies, we can approximate 
the conversion and emission zone as a thin slice (see also Ichiki et al. 2008). 
The electrons will have a completely cooled distribution, 
$N(\gamma)\propto \gamma^{-2}$, and the corresponding reprocessed 
emission will be characterised by a spectrum $F(\epsilon)\propto \epsilon^{-0.5}$. 
This power law extends up to an energy $\epsilon _{\rm max}$ determined by the maximum 
Lorentz factor of the produced pairs, in turn depending on $E_{\rm max}$, 
the maximum energy of the $\gamma$--rays emitted by the blazar, 
$\epsilon_{\rm max}\simeq 63 (E_{\rm max}/10 \,{\rm TeV})^2$ GeV.

The geometry we adopt is sketched in Fig. \ref{cartoon}. 
Since the emission from the blazar is strongly collimated into a cone with semi-aperture 
$\theta _{\rm c}\sim 1/\Gamma$, where $\Gamma$ is the bulk Lorentz 
factor of the jet, we can approximately assume that the source 
(located at the apex of the cone) uniformly illuminates the gray 
conical region in Fig. \ref{cartoon}.
Photons are converted into pairs that produce the GeV emission at the base of the cone.
 This geometry is different than that assumed in other works 
(e.g. Neronov \& Vovk 2010) that assume that the primary TeV emission is isotropic. 
In this case
the total flux recorded from the source does not depend on the intensity of the IGMF: 
only the surface brightness depends on it, since the 
effect of a larger IGMF is to broaden the (energy dependent) size of the reprocessed ``halo". 
Instead, in our scenario, the extension on the sky of the reprocessed emission  
cannot be larger than $\theta _{\rm v}\approx \theta_{\rm c}d_\gamma/(D-d_\gamma)$
where $d_\gamma$ is the distance (from the source) where the $\gamma$--$\gamma$ optical
depth is unity, and $D$ is the Earth--source distance.

The reprocessed emission contains all the absorbed flux. 
In absence of any magnetic field, $B=0$, the pairs are produced and emit 
along the direction of the primary $\gamma$--ray (Fig. \ref{cartoon}, upper panel) 
and the observer would measure a total reprocessed GeV flux equal to 
the absorbed TeV one (although in a different energy range). 
The condition that the absorbed flux is completely reprocessed as IC radiation is then:
\begin{equation}
F_{\rm abs} \equiv \int _{E_{\rm min}}^{E_{\rm max}} F_{\rm int}(E)[1-e^{-\tau(E)}]dE  =   
\int_{\epsilon_{\rm min}} ^{\epsilon _{\rm max}} F(\epsilon) d\epsilon 
\label{equalitynob}
\end{equation}
where $F_{\rm int}(E)$ is the source intrinsic spectrum (extending from 
$E_{\rm min}$ to $E_{\rm max}$) and $\tau(E)$ is the energy--dependent 
optical depth of $\gamma$--rays. 
Further assuming:
\begin{equation}
F(\epsilon)=k \left( \frac {\epsilon}{\epsilon _{\rm max}} \right)^{-0.5} 
\,\,\,\,\,\,\,\,\, \epsilon < \epsilon _{\rm max}
\label{specrepro}
\end{equation}
Eq. \ref{equalitynob} can  be used to derive the normalisation 
$k=F_{\rm abs}/(2\epsilon _{\rm max})$ (for $E_{\rm min} \ll E_{\rm max}$,  implying also $\epsilon _{\rm min} \ll \epsilon _{\rm max}$). 
We remark that, if the intrinsic blazar spectrum 
is harder than $E^{-1}$, the value of the absorbed flux $F_{\rm abs}$ 
(and, in turn, the level of the reprocessed radiation) depends solely on the maximum 
energy of the intrinsic emitted spectrum, $E_{\rm max}$.

The opposite case is that of a magnetic field so large that all the produced 
pairs are promptly isotropized in a time shorter than $t_{\rm cool}$ (bottom panel 
in Fig. \ref{cartoon}). 
The total flux recorded by the observer will be therefore strongly diluted. 
Note that in the isotropic case the observed 
reprocessed emission contains the contribution of {\it the two jets}, that  
pointing toward us and that pointing far from 
our line of sight. As above we can write:
\begin{equation}
\int_{\epsilon_{\rm min}} ^{\epsilon _{\rm max}} F(\epsilon) d\epsilon = 
2 \, F_{\rm abs} \frac{\Omega _{\rm c}}{4\pi}
\label{equalityiso}
\end{equation}
\noindent
where $\Omega _{\rm c}\simeq \pi \theta _{\rm c}^2$
is the solid angle in which the primary radiation is collimated. If the 
emitting source in moving with $\Gamma$ in one direction, then
$\Omega_{\rm c} \sim \pi/\Gamma^2$. The factor 2 takes into account the contribution of both jets.
The reprocessed spectrum is still given by Eq. \ref{specrepro} but with a 
normalisation $k=F_{\rm abs}\Omega _{\rm c}/(4\pi \epsilon _{\rm max})$.

\begin{figure*}
\vskip -2.4 true cm
\hskip -0.5 truecm
\hskip 1.1 truecm
\psfig{file=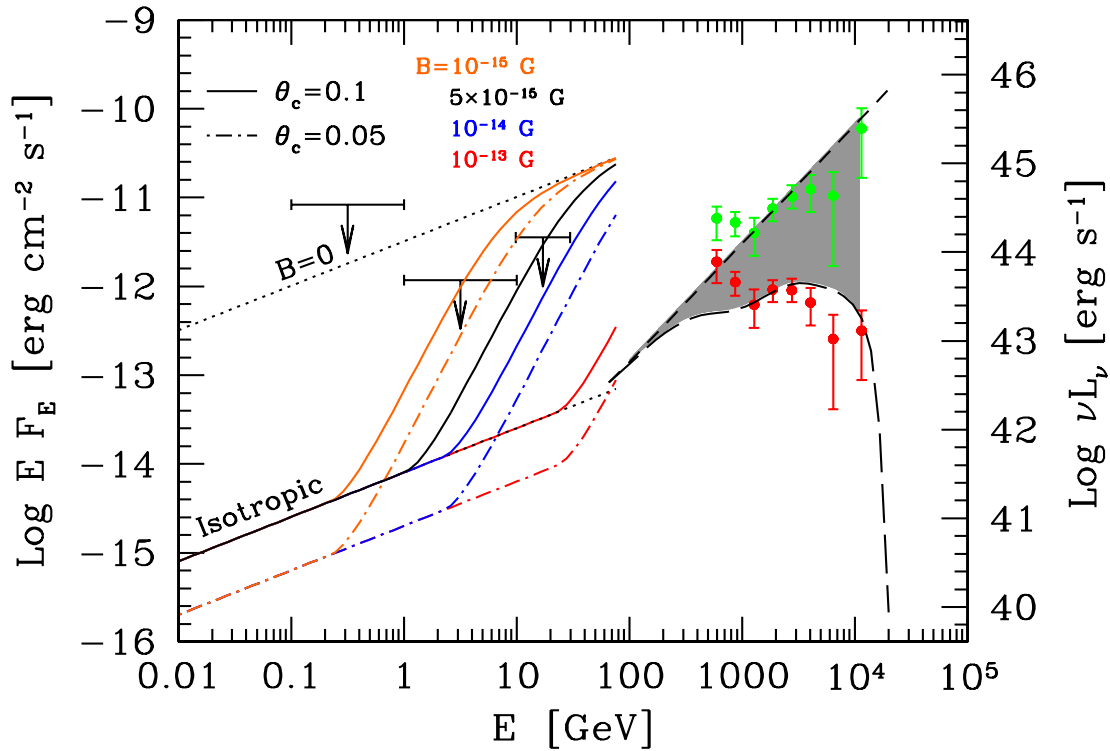,height=15.5cm,width=16cm} 
\vskip -3 true cm
\caption{
Spectral Energy Distribution of 1ES 0229+200 in the high 
energy band and the expected reprocessed GeV emission. 
Red points show the observed H.E.S.S. spectrum (Aharonian et al. 2007) 
and the green points the points after the correction for the absorption 
with the {\it LowSFR} EBL model of Kneiske et al. (2004). See Tavecchio et al. (2009) for details. 
The short-dashed black line is an approximation of the intrinsic spectrum, 
modelled as a hard power law. The long-dashed line is the corresponding 
absorbed spectrum following Kneiske et al. (2004). 
The shaded gray region between these two lines shows the absorbed flux. 
Note that the total amount of absorbed power, that is the relevant quantity for 
the estimate of the level of the reprocessed GeV emission, is only slightly 
sensitive on the assumed spectral shape, since it is dominated by 
the flux at the largest energies. 
The coloured lines report the approximation of the expected reprocessed 
spectrum for different values of the IGMF, $B=10^{-15}, 10^{-14}, 10^{-13}$ G 
and two different values of the initial collimation angle 
($\theta_{\rm c}=$ 0.1 rad and 0.05 rad, solid  and dashed--dotted 
lines, respectively), determining the intrinsic beaming of the primary radiation. 
For comparison, we also report the curves (black dotted lines) corresponding to 
the two extreme cases of (upper) $B=0$  and (lower) completely isotropy of the 
reprocessed emission (extremely large $B$).
Black points show the {\it Fermi}/LAT upper limit to the flux in the 100 MeV--1 GeV, 1--10 GeV and 10--30 GeV
bands obtained through the spectral parameters 
derived with the standard analysis (see text for more details). 
The solid black line shows the lowest possible reprocessed spectrum
consistent with the upper limits, corresponding to 
a magnetic field of $B=5\times 10^{-15}$ G (for $\theta _{\rm c}=0.1$).
}
\label{sed}
\end{figure*}

More complex is the intermediate case, for magnetic field such
that electrons with different Lorentz factors are deviated by different angles. 
In this case, the reprocessed emission is beamed into different solid angles 
$\Omega_{\gamma}=2\pi (1-\cos\theta _{\gamma})$ at different energies. 
The angle $\theta _{\rm \gamma}$ can be estimated assuming it is the angle 
by which the electron velocity vector changes in the cooling length $ct_{\rm cool}$:
%
$\theta _{\gamma} \sim ct_{\rm cool}/r_{\rm L} = 1.17\, B_{-15} (1+z_{\rm r})^{-4} \gamma_6^{-2}$
%
\noindent
where $r_{\rm L}=\gamma m_e c^2/(eB)\simeq 2\times 10^{24} \gamma _6 B _{-15}^{-1}$ 
cm is the Larmor radius of the electron. 

Since $\theta _{\gamma} \propto \gamma^{-2}$, we can derive the slope of the emitted spectrum in this regime:
\begin{equation}
F(\epsilon)d\epsilon \propto \frac {N(\gamma ) \, d\gamma \, 
\dot{\gamma}}{\theta _{\gamma}^2} \propto \gamma ^4 d\gamma  \;
\rightarrow \;
F(\epsilon) \propto \gamma ^4 \frac{d\gamma}{d\epsilon} \propto \epsilon ^{3/2}
\label{spectrum}
\end{equation}
where we used $N(\gamma)\propto \gamma ^{-2}$, $\dot{\gamma}\propto \gamma ^2$ 
(IC scattering), and $\epsilon\propto \gamma ^2$. 
Therefore the reprocessed 
spectrum in this region is described by a very hard power law 
(middle panel in Fig. \ref{cartoon}).

In real applications, there will be a combination of all the three cases 
described above: for a sufficiently low energy, 
$\epsilon < \epsilon _{\rm iso}$ the Lorentz factor of the corresponding emitting 
electron is sufficiently small to be isotropized by the magnetic field $B$ in 
its cooling time and the spectrum around $\epsilon$ will be thus given by 
Eq. \ref{equalityiso}. 
For energies above $\epsilon _{\rm iso}$ but lower than an 
energy $\epsilon _{\rm o}$ electrons curves substantially under the action of 
the magnetic field, but the emission is not isotropic and the spectrum is 
described by the hard power law $\epsilon ^{3/2}$. 
At energies above $\epsilon _{\rm o}$ (the energy at which approximately 
$\theta _{\gamma}\sim \theta _{\rm c}$), 
the reprocessed radiation is insensitive to the magnetic field, 
since the curvature of the electrons 
during their emission is (much) less than the beaming cone of the intrinsic 
blazar emission. 
The total spectrum will be described by three power laws 
and it can be well approximated by the following relation:
\begin{equation}
F(\epsilon)=k\left( \frac {\epsilon}{\epsilon _{\rm max}} \right)^{-0.5} 
\frac{1}{\Omega _{\rm c}+\Omega_{\gamma}}, \;\;\;\;\; \epsilon <\epsilon _{\rm max}.
\label{finale}
\end{equation}
%
The normalization $k$ is found in this case noting that  
Eq. \ref{finale} has to reduce to Eq. \ref{specrepro} for the case $B=0$. 
Inserting then $\Omega _{\rm \gamma}=0$ above and  comparing 
the resulting expression with Eq. \ref{specrepro}, we derive 
the value of the normalization: $k=\Omega _{\rm c} F_{\rm abs}/2\epsilon _{\rm max}$.

For simplicity, in  Eq. \ref{finale} we do not include the contribution from the receding jet, visible for energies at which the pairs are almost completely isotropised. 
This contribution would produce a doubling of the flux in the isotropic case, in a region of the spectrum not interesting for the case we are discussing here.

\section{Application to 1ES 0229+200}

\subsection{LAT data}

1ES 0229+200 is not present in the list of bright AGN from the first three 
months of Fermi Large Area Telescope (LAT) observations (Abdo et al. 2009). 
Therefore, we searched for detections in the publicly available 
data\footnote{\tt http://fermi.gsfc.nasa.gov}. We selected the photons of 
class 3 (diffuse) with energy in the range 0.1--100 GeV collected from 
2008 August 4 (MJD 54682) to 2010 March 31 (MJD 55286), for a total of 19 
months of elapsed time. 
These data were analyzed by using Science Tools 9.15.2, 
which includes the Galactic diffuse and isotropic background and the 
Instrument Response Function IRF P6 V3 DIFFUSE. 
After a selection of the useful events and good--time intervals, 
specifically taking into account a zenith angle $< 105$ deg to avoid 
the Earth albedo and the photons within the region of interest (ROI) 
with radius of 10 deg from the source radio position, we prepared a 
count map of the ROI. 
Then, we searched on the map for other possible contaminating sources inside 
the ROI, because we have to take into account them in the model. 
We model all the sources as single power laws, with flux and photon index free to vary.
The following steps are to calculate the live--time, the exposure map and 
the diffuse response. 
With all these information at hands, we performed an analysis by using an 
unbinned likelihood algorithm and calculated the corresponding test statistic 
($TS$, see Mattox et al. 1996 for a definition of $TS$; here it is worth 
noting that the significance sigma roughly equal to $\sqrt{TS}$ ). 
None of the energy bins shows a significance larger than $TS=25$, therefore only 
upper limits can be derived 
 (although for the 10--30 GeV bin there is a hint of detection at the
$\sqrt{TS}\sim 4\sigma$ level).
Following Abdo et al. (2010a), upper limits have been 
calculated looking at the flux satisfying $2\Delta \log (L)=4$  when increasing the flux
from the value minimizing the likelihood $L$.

The measured flux and photon index are the average over the 19 months of analysed data. 
Since the SED studied here are built with non simultaneous data and the 
1ES 0229+200 is a weak gamma--ray source, we have not studied the variability 
of the detected sources and used only the averages. 
The results are summarised in Table 1.
The latest available calibration of LAT indicates the presence of systematic 
errors, which have to be added to the quoted statistical errors in Table 1, 
and have value of 10\% at 100 MeV, 5\% at 500 MeV, and 20\% at 10 GeV 
(Rando et al. 2009).

\begin{table} 
\begin{center}
\begin{tabular}{lcccc}
\hline
\hline
 $E$ bin&UL&  N& $TS$\\
$\;$ (1)& (2)&  (3) & (4)\\
\hline  
0.1--1& 1.63$\times 10^{-8}$ & 155&  6.7\\ 
1--10& 2.34$\times 10^{-10}$  & 3& 1.55 \\
10--30&1.15$\times 10^{-10}$ & 2& 17.65 \\
\hline
\hline
\end{tabular}                                                         
\caption{Results of the analysis of the LAT data. (1) Energy band, in GeV. 
(2) Estimated upper limit (ph cm$^{-2}$ s$^{-1}$). (3): number of photons in the band. (4): 
value of $TS$. 
See text for details.}
\end{center}
\label{LAT}
\end{table}                                                                  

\subsection{Results}

Fig. 2 shows the high-energy spectral energy distribution of 
1ES 0229+200 including the TeV data from H.E.S.S. (red) and the same 
points corrected for the absorption (green) using the {\it Low SFR} model of Kneiske et al. (2004).  
We assume that the intrinsic spectrum is well represented 
(dashed line) by a hard power law, $F_{E}\propto E^{1/3} $ 
(see Tavecchio et al. 2009 for the justification of this choice). The black dotted line is the corresponding absorbed spectrum using the 
Kneiske et al. (2004) model. The area in gray show the flux absorbed and 
available for reprocessing. 
 As long as the intrinsic
spectrum is hard, the amount of absorbed energy depends only 
on the intrinsic luminosity of the highest energy bin $E_{\rm max}$. 
The most conservative limit on the IGMF corresponds to the lowest 
amount of reprocessed radiation that in turn corresponds to the smallest
intrinsic luminosity.
To this aim we use the EBL model providing the lowest opacity around 10 TeV.

We report the LAT upper limit in the 0.1--1 GeV, 1--10 GeV and 10--30 GeV band (Table 1). 
The solid and dot--dashed lines report the expected reprocessed 
emission assuming three different values of the IGMF and two 
different beaming angles (0.05 and 0.1 rad, corresponding to bulk Lorentz factor $\Gamma=20$ and 10, respectively) for the intrinsic blazar emission. 
The black line is calculated for the minimum value of the magnetic field consistent with the upper limit, $B=5\times 10^{-15}$ G. Note that, due to the very hard reprocessed spectrum, the 
most stringent upper limit is that at the highest energies, 10--30 GeV. 
Beaming angles $\theta _{\rm c}$ smaller than those assume here 
(corresponding to larger bulk Lorentz factors of the jet) would result 
in lower values for the upper limit on $B$, see Eq. \ref{finale}.

We remark that, unlike the  
methods based on the estimate of the rotation measure in the 
radio band (e.g. Kronberg 2001), with which it is possible to derive upper limits to the 
IGMF, this methods allows us to put a {\it lower limit} on $B$.
If the hint of detection in the highest energy bin is real, we have two possibilities:
either it is the reprocessed radiation, and in this case it gives a {\it measure} of $B$,
or it is still primary emission from the blazar (even if belonging to a  
component different than that observed at TeV energies, e.g. Tavecchio et al. 2009),
and in this case our limit would still hold.

\section{Discussion}

The lower limit on the value of the magnetic field derived here, 
$B>5\times 10^{-15}$ G can be considered one of the most stringent value 
ever derived for the IGMF. 
The value is mainly constrained by the 
LAT upper limit above 10 GeV, in which the source is tentatively detected 
at the 4$\sigma$ level. 
We remark that, since we derive a lower 
limit on $B$, even if this {\it Fermi}/LAT measure is considered 
as an upper limit the conclusion does not change. 

The main assumptions adopted in this paper are: 
(i): the amount of reprocessed energy is derived from the level 
of the observed spectrum measured by H.E.S.S. (Aharonian et al. 2007); 
(ii) we use the {\it Low SFR} Kneiske et al. (2004) model to calculate the optical depth; 
(iii) we assume that the maximum energy of the intrinsic absorption is 11 TeV. 

Assumption (i) is supported by the extremely small variability observed by H.E.S.S. during the observations in 2005--2006. 
Of course, any variations of the intrinsic emission of the blazars, will be reflected into the reprocessed emission. Ideally, simultaneous 
TeV and GeV observations would be required to take into account variability and exploit the information carried by the delay 
between the intrinsic TeV and the reprocessed GeV emission (see Dai et al. 2002, Murase et al. 2008 for discussions). However, this approach is prevented in the specific case of 1ES 0229+200 by the very small GeV flux, still at the detection limit of LAT after 18 months of observations. Also in other TeV BL Lacs more bright in the TeV band, integration over few days are necessary to detect the GeV emission (e.g. Abdo et al. 2010b). Moreover, in these objects, the intrinsic GeV emission is much more luminous than the expected reprocessed radiation. These features make difficult to reveal the reprocessed radiation even during bright flares. 
 
The use of a specific model for the absorption is unavoidable. 
However, several of recent EBL models converge at the wavelengths above 
10 $\mu$m, the ones determining the opacity for $\gamma$--rays of energy below 10 TeV. 
In particular, the recent models of 
Franceschini et al. (2008), Gilmore et al. (2009), Finke et al. (2010) agree well (see e.g. 
the discussion of Finke et al. 2010) and are consistent with the low level of the EBL suggested 
by recent observations of Cherenkov telescopes (e.g. Aharonian et al. 2006, Mazin \& Raue 2007; 
see also Kneiske \& Dole 2010).
The model adopted here ({\it LowSFR} of Kneiske et al. 2004) provides an optical depth similar to 
that of all the other updated models up to energies of 4--5 TeV. For larger energies the predicted optical depth is {\it lower} than that of the other models (see also Tavecchio et al. 2009). 
We again stress that this implies to minimize the luminosity of the reprocessed radiation, and hence 
the derived lower limit of $B$ is a conservative value.

The assumed maximum energy of the intrinsic emission, point (iii), is somewhat critical, since it determines the total amount of energy reprocessed and then re-emitted in the GeV band, and the possible development of cascades. The first point is clear: since the intrinsic spectrum is rather hard, the total luminosity of reprocessed radiation (and its maximum energy) depends on $E_{\rm max}$. Under the assumption that the spectrum is proportional to $E^{1/3}$, the normalization of the reprocessed emission is (see Eq. \ref{finale}) $k \propto E_{\rm max}^{4/3}/\epsilon_{\rm max}\propto E_{\rm max}^{2/3}$. Therefore all the curves reported in Fig.\ref{sed} will shift upward by this factor. We consider unlikely that the spectrum extends unbroken at energies well above few tens of TeV and therefore the correction factor cannot be much larger than a factor of a few. In any case we remark again that, since  a larger $E_{\rm max}$ would have the effect to {\it increase} the value of the derived lower limit (a larger $B$ would be required to downshift the curves), our limit can be considered also in this respect, again, as a conservative value. The same argument applies when considering the possible role of electromagnetic cascades. For values of $E_{\rm max}$ larger than $\sim 30$ TeV a sizeable amount of the reprocessed flux would be emitted above $\sim 500$ GeV (the energy above which the optical depth for pair conversion exceeds 1) and then reabsorbed, in turn producing new pairs and thus initiating a cascade. In this case, apart the different resulting spectrum, the total flux of the reprocessed component would be larger than what simply estimated here: therefore, also in this case, our value of $B$ can considered a conservative value.
 
Finally, we note that in our derivation we are implicitly assuming that the magnetic field is 
oriented perpendicularly to the direction of the relativistic pairs. In reality the IGMF is probably randomly oriented, maintaining its coherence in domains with a size $\sim$1 Mpc (e.g. 
Neronov \& Semikoz 2009). Including the geometry of the field in our calculations would result 
in a limit to the IGMF slightly larger than that derived here. 
Again, our inferred value of $B$ is then a conservative one.

\section*{Acknowledgements}
We thank the referee for helpful comments.
We are grateful to L. Costamante for useful comments and discussions.
This work was partly financially supported by a 2007 COFIN-MiUR grant.

\end{document}